# Cluster superconductivity in the magnetoelectric Pb(Fe$_{1/2}$Sb$_{1/2}$)O$_3$ ceramics

V.V. Laguta[1,2], M. Marysko[1], V. A. Stephanovich[2], I. P. Raevsky[3], N.M. Olekhnovich[4], A.V. Pushkarev[4], Yu.V. Radyush[4], S.I. Raevskaya[3], R.O. Kuzian[5], V. Chlan[6], H. Štěpánková[6]

[1]*Institute of Physics AS CR, Cukrovarnicka 10, 162 53 Prague, Czech Republic*
[2]*Institute of Physics, Opole University, Oleska 48, 45-052 Opole, Poland*
[3]*Research Institute of Physics and Faculty of Physics, Southern Federal University, Stachki Ave. 194, 344090, Rostov-on-Don, Russia*
[4]*Scientific-Practical Materials Research Centre, NASB, Minsk, Belarus*
[5]*Institute for Problems of Materials Science, National Academy of Sciences of Ukraine, Krjijanovskogo 3, 03142 Kyiv, Ukraine*
[6]*Charles University in Prague, Faculty of Mathematics and Physics, Prague 1, Czech Republic*

**Abstract** We report the observation of cluster (local) superconductivity in the magnetoelectric Pb(Fe$_{1/2}$Sb$_{1/2}$)O$_3$ ceramics prepared at a hydrostatic pressure of 6 GPa and temperatures 1200-1800 K to stabilize the perovskite phase. The superconductivity is manifested by an abrupt drop of the magnetic susceptibility at the critical temperature $T_C \approx 7$ K. Both the magnitude of this drop and $T_C$ decrease with magnetic field increase. Similarly, the low-field paramagnetic absorption measured by EPR spectrometer drops significantly below $T_C$ as well. The observed effects and their critical magnetic field dependence are interpreted as manifestation of the superconductivity and Meissner effect in metallic Pb nanoclusters existing in the ceramics. Their volume fraction and average size were estimated as 0.1-0.2% and 140-150 nm, respectively. The superconductivity related effects disappear after oxidizing annealing of the ceramics.

**I. Introduction**

Modern technologies permit to fabricate the materials and interfaces with ordinary crystal structure but unusual physical properties combining, for instance, long range magnetic and ferroelectric orders with local superconductivity. Well-known examples are the interfaces between perovskite oxides [1,2], which may have all the above properties [1,3,4] along with metallic interfacial conductivity [5-8]. Here we report the first time observation of local superconductivity in the magnetoelectric Pb(Fe$_{1/2}$Sb$_{1/2}$)O$_3$ (PFS). This material possesses both magnetism and ferroelectricity like other much more known chemically disordered double perovskites PbFe$_{1/2}$Nb$_{1/2}$O$_3$ (PFN) and PbFe$_{1/2}$Ta$_{1/2}$O$_3$ (PFT) [9-17]. PFS shows quite intriguing magnetic properties such as existence of dynamic magnetic nanoregions with large frustrated magnetic superspins, which on cooling freeze in superspin glass state coexisting with the long-range ordered antiferromagnetic (AFM) phase at T < 32 K [9]. Such behavior of PFS is drastically different from that known for its disordered counterparts PFN and PFT where the long-range AFM ordered phase exists below $T_N$ = 150-155 K [12,18-20], while below $T_g$ = 11-12 K it coexists with the spin glass phase [11,19,20]. This difference was attributed to the fact that PFS is highly ordered [9]. Recently, we were able to fabricate, for the first time, the PFS ceramics with different degree of long-range chemical ordering between the magnetic $Fe^{3+}$ and non-magnetic $Sb^{3+}$ ions seen well by X-ray diffraction. This fabrication flexibility permits to synthesize the samples with different chemical compositions and thus to clarify the role of chemical order/disorder in magnetic properties of double perovskites.

At room temperature, chemically ordered PFS has a simple cubic perovskite structure with space group $Fm\bar{3}m$. On cooling to 200 K, the cubic structure transforms to polar one as evidenced from the dielectric permittivity and hysteresis loops data [9]. Its X-ray diffraction data have been reported in Ref. [21] where the crystal structure of the low-temperature polar phase was not refined yet. Other magnetic and structure data obtained from *ab initio* Density functional theory (DFT) calculations are presented in Refs. [9,22,23].

In this paper we pay attention mainly to unusual behavior of PFS ceramics at temperatures below 10 K, namely abrupt decrease of the magnetic susceptibility along with appearance of low-field EPR absorption and its critical dependence on applied magnetic field, which is interpreted as a manifestation of cluster (local) superconductivity due to presence of metallic Pb clusters in ceramics. Besides, as the measurements were performed on samples ranging from nearly ordered (s=0.93) to almost disordered (s=0.21), we observed the transformation of magnetic structure with the change of chemical ordering. In particular, the ground magnetic state of the disordered sample contains only spin glass phase with the freezing temperature $T_g$ = 25 K in contrast to disordered PFN and PFT, where the long-range AFM ordered phase exists below $T_N$ = 150-155 K [12,18-20].

**II. Experimental**

PFS ceramic samples were prepared in two stages [21]. First, we synthesized the stoichiometric composition $Pb(Fe_{1/2}Sb_{1/2})O_3$ from the initial PbO, $Fe_2O_3$ and $Sb_2O_5$ oxides at T= 1020–1030K for 4 hours. The resulting compound had a pyrochlore crystal structure. The second stage of this synthesis has been performed under hydrostatic pressure 6 GPa and temperature 1200 -1800 K for 2-10 min. After the synthesis, the system was rapidly cooled to the room temperature under pressure and only then the high-pressure apparatus was unloaded. The product of the second high-pressure synthesis was dense coarse-grained (the grain size varied from 1 to 5 μm with a mean size of approximately 2 μm) PFS ceramics with single phase perovskite structure. Room temperature x-ray diffraction patterns of PFS correspond to a cubic symmetry (space group Fm3m [21]) and exhibit superstructure lines attributed to a double perovskite unit cell caused by partial chemical 1:1 ordering of $Fe^{3+}$ and $Sb^{5+}$ ions. The mean value of the chemical ordering degree varies from s=0.17 up to s=0.93 depending on synthesis conditions. It is determined either from the ratio of the intensities of the superstructure XRD reflections to the fundamental ones, or the fraction of a singlet component in Mossbauer spectrum [24]. The samples with s=0.67 and 0.21 were annealed in air at T≈770 K in order to oxidize residual Pb metallic inclusions in them.

The magnetic measurements were carried out using the SQUID magnetometer MPMS-5S (Quantum Design) under several important protocols including field cooling (FC) and zero-field cooling (ZFC) in the dc regime. Electron paramagnetic resonance (EPR) measurements were performed at 9.407 GHz in a temperature range from 3.5 to 300 K, by employing the Bruker E580 spectrometer and Oxford Instrument cryostat.

**III. Results**

Fig. 1 presents magnetic susceptibility data measured at ZFC and FC regimes in the field 500 Oe for few PFS samples with different degree of chemical ordering between Fe and Sb ions from s = 0.21 to s = 0.93. The temperature behavior of the susceptibility in the sample with high degree of chemical ordering is identical to that described in our previous paper [9]. On cooling, susceptibility exhibits an abrupt increase at approximately 250 K due to the formation of superparamagnetic phase or superspins. Then, in the temperature range between 100 and 150 K, the ZFC magnetic susceptibility (dashed lines) shows a broad maximum related to collective freezing into a superspin glass phase. And finally, the second, sharp maximum in both FC and ZFC data characterizes AFM phase transition with the Neel temperature $T_N$ = 32 K. One can see that with the chemical order decrease, the anomalies related to the superspin glass phase gradually disappear. Likewise, the ZFC peak at ~ 30 K transforms into a cusp characterizing transition to classical spin glass state where ZFC and FC curves essentially differ below the peak temperature. However, in this paper we shall concentrate on the abrupt decrease of susceptibility on further cooling down to T < 7 K visible in all three samples (Fig. 1a).

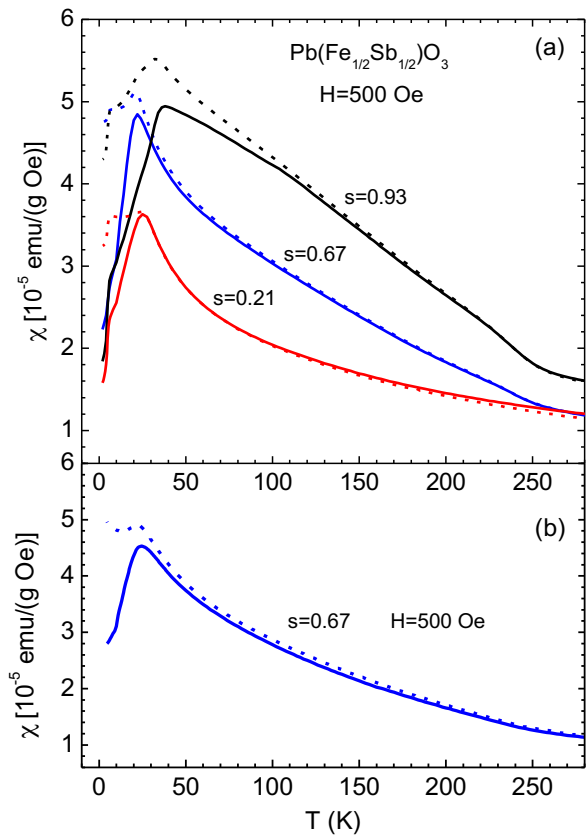

**Fig. 1.** (a) Temperature dependence of the FC (dashed lines) and ZFC (solid lines) magnetic susceptibilities for three PFS samples having different degrees of the chemical ordering: s = 0.93; 0.67; and 0.21. (b) Magnetic susceptibility versus temperature for the sample with s=0.67 after its annealing in air at T≈770 K.

Note that this abrupt decrease of the susceptibility disappears after annealing of samples in air at approximately 770 K as it is shown in Fig. 1b for the sample with s=0.67. Since the anomaly at T<7 K is observed for all three samples, below we report the detailed studies of only the sample with s = 0.21 for which the susceptibility does not expose to the influence of superspin glass phase.

Fig. 2 reports magnetic susceptibility for this sample measured at different magnetic fields from 50 to 2000 Oe. The abrupt decrease of susceptibility at $T_C \approx 7$ K, which is much lower than $T_g \approx 23$ K, is well seen in both FC and ZFC curves. However, the $T_C$ starts to decrease with field increase and this anomaly completely disappears for the fields above of only 1000 Oe.

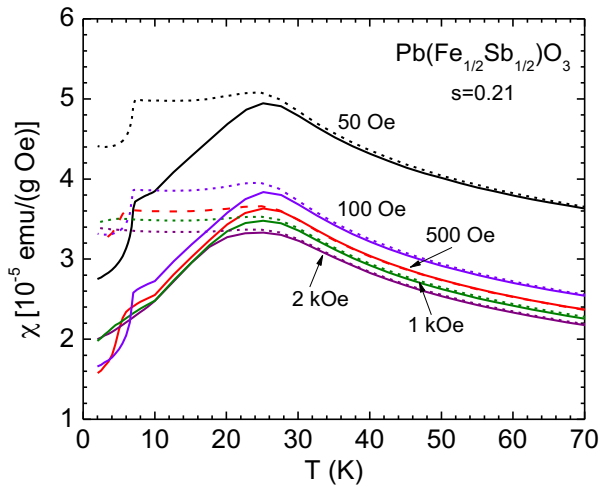

**Fig. 2.** Temperature dependence of the FC (dashed lines) and ZFC (solid lines) magnetic susceptibilities of PFS sample with s = 0.21 at different magnetic fields.

We have measured also EPR spectra for these PFS samples. As an example, Fig. 3 shows EPR spectra for the sample with s = 0.21 measured in the temperature range from 7.3 down to 3.7 K. One can see that at T < 7 K, a new line appears at low magnetic fields. This line shifts towards higher magnetic fields with further sample cooling and has nontypical line shape, which does not resemble usual first derivative of the Lorentzian or Gaussian curves. Note that EPR spectrometer measures first derivative of EPR absorption signal due to the use of a lock-in detection system. This line disappears after sample annealing in air (Fig. 3b) similar to the anomaly in magnetic susceptibility at T<7 K.

The unusual line shape indicates that the EPR absorption decreases abruptly at the resonance peak field. This is well seen in the absorption spectra obtained by the integration of the virgin one (Fig. 3c). One can also observe the shift of the critical field related to the observed anomaly towards higher magnetic fields with temperature lowering. Such low-field EPR signal is often observed in superconductors (see, e.g. Refs. 25-27) due to change in the diamagnetic susceptibility upon transition from the Meissner (i.e. superconductive) state [28] to the mixed or normal (i.e. non-superconductive) one. Therefore, both magnetic susceptibility and EPR data suggest that the observed phenomenon may be related to superconductivity in some regions of our samples. Defining the critical temperature $T_c$ as the onset of the negative diamagnetic contribution to susceptibility and the critical field $H_c$ as the value of the magnetic field at which the diamagnetic contribution disappears, we observe that the corresponding values $T_c \approx 7$ K and $H_c \approx 700$ Oe (at 3.7 K) coincide approximately with those of superconducting Pb [29-30]. This is not much surprising as our samples were fabricated on the base of PbO oxide. Due to specific synthesis condition (high pressure and rapid cooling) Pb can be partially reduced, for instance, at grain boundaries. It can be again oxidized by annealing in air. PFS ceramics after the oxidizing annealing does not show any anomalies at T<7 K in both magnetic susceptibility and EPR spectra (see, Figs. 1b and 3b).

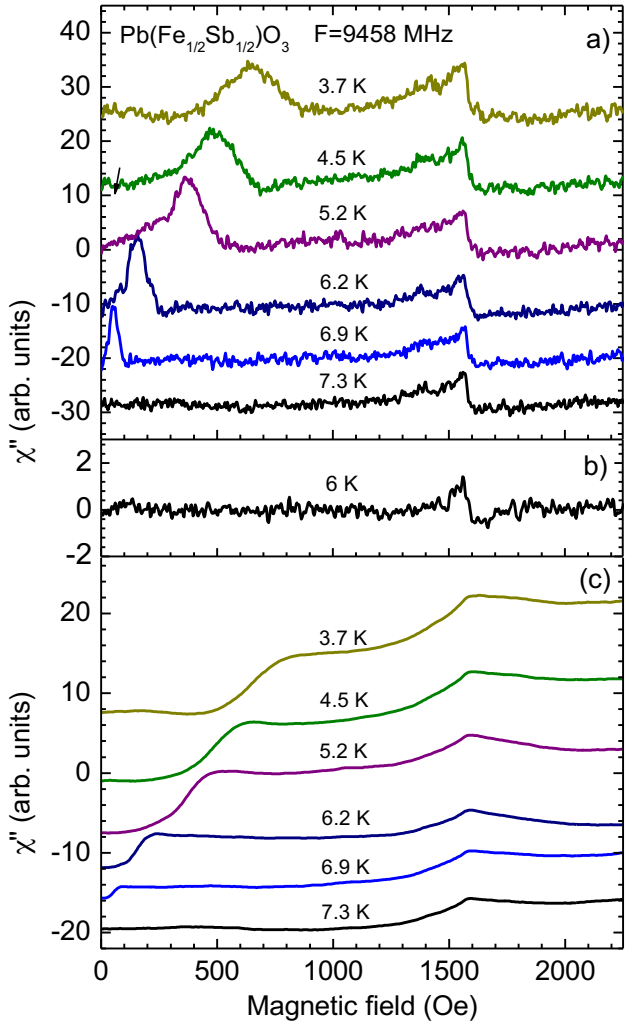

**Fig. 3.** EPR spectra measured in PFS (s = 0.21) at temperatures from 7.3 down to 3.7 K. Panels (a) and (b) report the first derivative of EPR absorption signal in the virgin (a) and annealed in air (b) samples. Integrated absorption spectra are shown in panel (c).

To verify origin of the observed EPR signal, we measured, as a test, EPR spectra of small Pb metallic particles (~0.1 mm) mixed with epoxy resin. One can see (Fig. 4) that below the critical temperature $T_c \approx 7.2$ K, low-field spectral line appears similar as in PFS ceramics. The magnetic field value at which the EPR absorption peak occurs is designated as the critical field $H_c$ [25-27]. The temperature dependence of this critical field for Pb particles and PFS ceramics is reported in Fig. 5. One can see that our PFS ceramics data are in good coincidence with those in bulk testing sample and literature data [29-30] except the lowest temperatures where PFS exhibits higher critical fields due to size effects, i.e. the increase of the critical field in small particles [29-30].

It is worth also to mention that the spectra similar to those in Figs. 3 and 4 were reported for small (0.1-1 mm) superconductive thin spheres [31].

The above experimental facts show convincingly that the observed phenomenon in PFS ceramics is related to superconductivity of local Pb clusters in it. Such clusters can emerge at the grain boundaries. The full screening of some grains by Pb shell cannot be excluded as well. However these inclusions of Pb do not form the percolative cluster as the room-temperature ac conductivity of all the samples studied is rather low ($10^{-8}$ - $10^{-7}$ S/cm at 1 kHz and $10^{-5}$ - $10^{-4}$ S/cm at 1 MHz). Annealing of the samples in air at 700-800 K completely destroys the superconductivity as the metallic Pb inclusions transform into PbO due to oxidation. It is worth noting that $^{207}$Pb NMR measurements also show presence of Pb ions which do not belong to intrinsic magnetic composition.

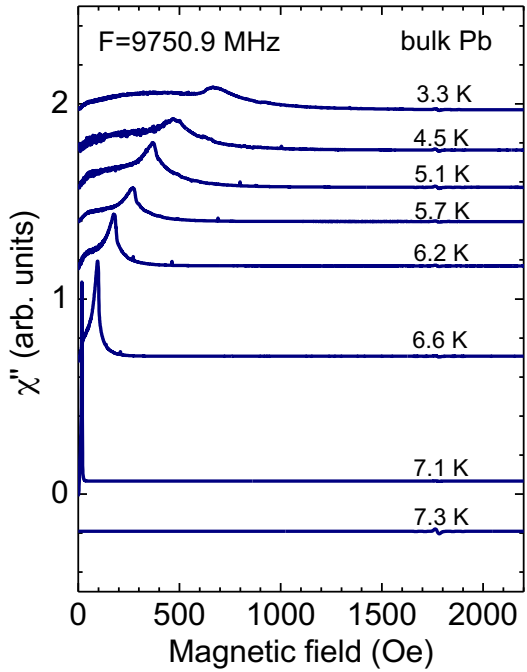

**Fig. 4.** EPR spectra measured in bulk metallic Pb sample.

To assess the volume of Pb clusters, $V_{Pb}$, in our PFS sample (for definiteness we choose s=0.21), we use the results of the work [29] dealing with the superconductivity of Pb nanoparticles. In this work, the experimental magnetic field dependence of superconducting critical temperature $T_c$ has been approximated by the relation [29]:

$$H_C(T) = H_C(0)\left\{1 - \left(\frac{T_C}{T_C(0)}\right)^\alpha\right\}, \quad (1)$$

where $T_C(0)$ and $H_C(0)$ are, respectively, the critical temperature at H=0 and critical magnetic field at T=0. The exponent $\alpha$ is related to the characteristics of magnetic field penetration into a superconductor and increases from 2 (bulk Pb), to 3 for the nanoparticles with diameter 6 nm [29]. We consider the value of $T_C$=7.2 K to be equal to $T_C(0)$. As our $T_C(0)$ coincides with that of bulk Pb, it is reasonable to put $\alpha$=2 in subsequent calculations. The best fit of the above expression to the data in the Fig. 5 (solid line) yields $H_C(0) \approx 950$ Oe. The estimate of linear dimension *d* of a Pb cluster can be done with the empirical relation (see Fig. 8 of Ref. 29) $H_C(0) = 170 d^{-3/2}$ (where *d* is in nanometers and $H_C(0)$ is in Tesla) from which we obtain $d \approx 140\text{-}150$ nm.

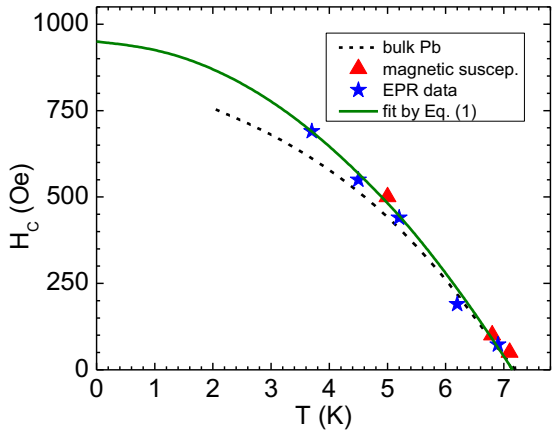

**Fig. 5.** Critical field versus temperature in bulk Pb (dashed line) and PFS ceramics determined from magnetic susceptibility (triangles) and EPR spectra (stars). The solid line is the fit to Eq. (1).

To estimate the volume portion of metallic Pb in our PFS sample, we use the simple argument stemming from the low-temperature behavior of magnetic susceptibility in Fig. 2. Namely, ZFC susceptibility in the low filed H=50 Oe decreases by $\Delta\chi_m \approx 9.8 \cdot 10^{-6}$ emu/(g Oe) on cooling from 7.2 K down to 2 K. This mass susceptibility decrease corresponds to the volume value $\Delta\chi_V \approx 9 \cdot 10^{-5}$ emu/(cm$^3$ Oe). Using this value we can estimate the entire volume of Pb fraction that becomes superconductive, as approximately $\Delta\chi_V / \chi_0 \approx 0.0012$ where $\chi_0 = -1/4\pi$ is the volume susceptibility of a bulk superconductor. Taking into account penetration depth of magnetic flux into Pb cluster (~40 nm), the total volume of metallic Pb clusters may be few times larger than the above estimated value 0.12%. Such minute amount of Pb clusters randomly distributed in ceramics cannot be detected by transmission electron microscopy. However, they are reliably detectable via observation of their superconductive properties in magnetic susceptibility and magnetic resonance spectra even in a quantity as low as few micrograms.

Discovered local superconductivity of PFS ceramics may take place in other Pb contained complex oxide materials, including many double perovskites (PFN, PFT, Pb(Fe$_{1/2}$W$_{1/2}$)O$_3$ (PFW)) as well as PZT and PMN-PT ceramics. In particular, we detect the metallic Pb inclusions in PFT ceramics synthesized with 6% excess of PbO oxide [32]. Such ceramics exhibits properties similar to described above. Inclusions of metallic Pb have been already revealed in PbVO$_3$ [33] and Ba$_{1-x}$La$_x$PbO$_3$ [34] ceramics obtained by high-pressure synthesis. In the latter material, the anomaly of magnetic susceptibility at about 7 K has also been observed. It was attributed to a transition of Pb inclusions into the superconducting state, though this anomaly was masked by the transition into the superconducting state of the Ba$_{1-x}$La$_x$PbO$_3$ phase, which occurs at somewhat higher temperature. Of course, this effect will not influence markedly the electric or magnetic properties of a material at temperatures above $T_C \approx 7$ K. But it will have impact below the critical temperature leading to unusual behavior of magnetic and dielectric characteristics, which could be misinterpreted. For instance, the abrupt decrease of the magnetic or dielectric susceptibility and relaxation at $T_C$ can be interpreted as a manifestation of phase transition or even quantum effects, like the quantum spin tunneling [35]. Strong influence of the external magnetic field on susceptibility in Pb contained magnetoelectrics at low temperatures can be wrongly interpreted as manifestation of magnetoelectric coupling as well.

In summary, we have synthesized magnetoelectric $Pb(Fe_{1/2}Sb_{1/2})O_3$ ceramics with controlled degree of chemical ordering ranging from s=0.17 (almost disordered) up to s=0.93 (nearly perfectly ordered). In contrast to chemically disordered $Pb(Fe_{1/2}Nb_{1/2})O_3$ or $Pb(Fe_{1/2}Ta_{1/2})O_3$, the disordered PFS ceramics has a unique magnetic ground state in the form of short-range ordered spin-glass phase with the freezing temperature around 25 K. We have additionally found that above PFS ceramics shows quite unusual behavior of magnetic susceptibility and EPR signal. Namely, both these quantities sizeably drop at the critical temperature ≈7 K, which depends on applied magnetic field. We explained these observations as manifestation of local superconductivity of Pb metallic clusters at grain boundaries. Estimated total volume of these metallic clusters is of the order $(1-2) \cdot 10^{-3}$ of the sample volume and its typical linear dimension is around 140-150 nm.

ACKNOWLEDGMENTS

The research was supported by the GA CR under project No. 13-11473S, projects SAFMAT LM2015088 and LO1409 and partially by the Russian Foundation for Basic Research (grant T16R-079), Belarusian Republican Foundation for Fundamental Researches (grant 16-52-00072 Bel_a), the Ministry of education and science of the Russian Federation (research project 2132)